\documentclass[aps,showpacs,preprintnumbers,amsmath,amssymb,nofootinbib,preprint,showkeys]{revtex4}

\usepackage{graphicx}
\usepackage{color}
      
\def \be  {\begin{equation}}
\def \ee  {\end{equation}}
\def \ee  {\end{equation}}
\def \bea {\begin{eqnarray}}
\def \eea {\end{eqnarray}}

\def \Tr  {\bf{Tr}}

\begin{document}

\preprint{ECTP-2013-05}

\title{Chemical Freeze-Out and Higher Order Multiplicity Moments}

\author{A.~Tawfik}
\email{a.tawfik@eng.mti.edu.eg}
\email{atawfik@cern.ch}
\affiliation{Egyptian Center for Theoretical Physics (ECTP), MTI University, Cairo, Egypt}
\affiliation{World Laboratory for Cosmology And Particle Physics (WLCAPP), Cairo, Egypt}

\date{\today}

\begin{abstract}
We calculate the non-normalized moments of the particle multiplicity within the framework of the hadron resonance gas (HRG) model. At finite chemical potential $\mu$, a non-monotonic behavior  is observed in the thermal evolution of third order moment (skewness $S$) and the higher order ones as well. Among others, this observation likely reflects dynamical fluctuations and strong correlations. The signatures of non-monotonicity in the normalized fourth order moment (kurtosis $\kappa$) and its products get very clear. Based on these findings, we introduce a novel condition characterizing the universal freeze-out curve. The chemical freeze-out parameters $T$ and $\mu$ are described by vanishing $\kappa\, \sigma^2$ or equivalently $m_4=3\,\chi^2$, where $\sigma$, $\chi$ and $m_4$ are the standard deviation, susceptibility and fourth order moment, respectively. The fact that the HRG model is not able to release information about criticality related to the confinement and chiral dynamics should not veil the observations related to the chemical freeze-out. Recent lattice QCD studies strongly advocate the main conclusion of the present paper.

\end{abstract}


\maketitle


\section{Introduction}

It is conjectured that the partition function of the hadron resonance gas (HRG) gives a satisfactorily approximation to the non-singular part of the free energy of the Quantum Choromodynamics (QCD), especially in the hadronic phase. The higher order  moments are assumed to reflect the large fluctuations associating the hadron-quark phase transition \cite{Tawfik:2012si}. This was the motivation of a remarkable number of experimental and theoretical studies \cite{endp1,star1,star2,goore}. Recently, various calculations have shown that the higher order moments of the multiplicity distributions of some conserved quantities, such as net-baryon, net-charge, and net-strangeness, are sensitive to the correlation length $\xi$ \cite{endp4,endp4a,corrlenght}, which in turn is related to the higher order moments, themselves. In realistic heavy-ion collisions, the correlation length is found to remain finite.  Furthermore, the higher order multiplicity moments have gained prominence in high energy physics with a huge hope in pinpointing the QCD critical endpoint (CEP) \cite{endp1} that likely connects the first order boundary separating the hadronic from the partonic matter at high density with the cross-over boundary at  low density \cite{endp2,Tawfik:2004sw}.  In Ref \cite{Tawfik:2012si}, a list of different moments of the particle multiplicity is introduced.  As we move from lower order to higher order multiplicities, certain distribution functions are being added / subtracted. These distribution functions represent higher order correlations \cite{Tawfik:2012si}. 

The particle multiplicities and their correlations are the observables providing information about the nature, composition and size of the medium from which they are originating. To determine the freeze-out parameters at various center-of-mass energies $\sqrt{s_{NN}}$, we utilize the analysis of the particle yields in terms of temperature $T$ and baryon chemical potential $\mu$. The baryon density in the system which is related to the chemical potential is generated by the nucleon stopping in the collision region. The chemical freeze-out is defined as the stage in the evolution of the hadronic system when inelastic collisions cease and the relative particle ratios become fixed. Both $T$ and $\mu$ can be related to $\sqrt{s_{NN}}$ \cite{jean2006}.

Many reasons speak for utilizing the HRG model in predicting the hadron abundances and their thermodynamics. The HRG model seems to provide a good description for the thermal  evolution of the thermodynamic quantities in the hadronic matter~\cite{Tawfik:2004sw,Karsch:2003vd,Karsch:2003zq,Redlich:2004gp,Tawfik:2004vv,Tawfik:2006yq,Tawfik:2010uh,Tawfik:2010pt,Tawfik:2012zz} and has been successfully utilized to characterize the conditions deriving the chemical freeze-out at finite densities~\cite{Tawfik:2005qn,Tawfik:2004ss,Tawfik:2012si}. In light of this, HRG can be used in calculating the higher order  moments of particle multiplicity using a grand canonical partition function of an ideal gas with all experimentally observed states up to a certain large mass as constituents. The HRG grand canonical ensemble includes two important features \cite{Tawfik:2004sw}; the kinetic energies and the summation over all degrees of freedom and energies of the resonances. On other hand, it is known that the formation of resonances can only be achieved through strong interactions~\cite{Hagedorn:1965st}; {\it Resonances (fireballs) are composed of further resonances (fireballs), which in turn consist of  resonances (fireballs) and so on}. In other words, the contributions of the hadron resonances to the partition function are the same as that of free particles with some effective mass. At temperatures comparable to the resonance half-width, the effective mass approaches the physical one \cite{Tawfik:2004sw}. Thus, at high temperatures, the strong interactions are conjectured to be taken into consideration through including heavy resonances. It is found that the hadron resonances with masses up to $2\;$GeV are representing suitable constituents for the partition function ~\cite{Karsch:2003vd,Karsch:2003zq,Redlich:2004gp,Tawfik:2004sw,Tawfik:2004vv,Tawfik:2006yq,Tawfik:2010uh,Tawfik:2010pt,Tawfik:2012zz}. Such a way, the singularity expected at the Hagedorn temperature~\cite{Karsch:2003zq,Karsch:2003vd} can be avoided and the strong interactions are assumed to be considered. Nevertheless, the validity of HRG is limited to temperatures below the critical one, $T_c$.

In this paper, we study the non-normalized higher order moments of the particle multiplicity in  the HRG model in section \ref{sec:model}. The normalized higher order moments are discussed in section \ref{sec:norm}. The thermal evolution of second, third and fourth order moments and their products (ratios) are studied at different chemical potentials $\mu$. Section \ref{sec:chemFO} is devoted to introduce the novel condition describing the freeze-out parameters and their dependence on $\sqrt{s_{NN}}$. The conclusions are outlined in section \ref{sec:conc}.


\section{The Hadron Resonance Gas Model}
\label{sec:model}

In a grand canonical ensemble, it is straightforward to derive an expression for the pressure.  The hadron resonances treated as a free gas~\cite{Karsch:2003vd,Karsch:2003zq,Redlich:2004gp,Tawfik:2004sw,Tawfik:2004vv} are conjectured to add to the thermodynamic pressure in the hadronic phase (below $T_c$). This statement is valid for free as well as for strongly interacting resonances. 
It has been shown that the thermodynamics of strongly interacting  system can also be approximated to an ideal gas composed of hadron resonances with masses $\le 2~$GeV ~\cite{Tawfik:2004sw,Vunog}. Therefore, the confined phase of QCD, the hadronic phase, is modelled as a non-interacting gas of resonances. The grand canonical partition function reads
\bea
Z(T, \mu, V) &=&\Tr\left[ \exp^{\frac{\mu\, N-H}{T}}\right]
\eea
where $H$ is the Hamiltonian of the system and $T$ ($\mu$) is the temperature (chemical potential). The Hamiltonian is given by the sum of the kinetic energies of relativistic Fermi and Bose particles. The main motivation of using this Hamiltonian is that it contains all relevant degrees of freedom of confined and  strongly interacting matter. It includes implicitly the interactions that result in resonance formation. In addition, it has been shown that this model can submit  a quite satisfactory description of particle production in heavy-ion collisions. With the above assumptions the dynamics the partition function can be calculated exactly and be expressed as a sum over 
{\it single-particle partition} functions $Z_i^1$ of all hadrons and their resonances.
\bea \label{eq:lnz1}
\ln Z(T, \mu_i ,V)&=&\sum_i \ln Z^1_i(T,V)=\sum_i\pm \frac{V g_i}{2\pi^2}\int_0^{\infty} k^2 dk \ln\left\{1\pm \exp[(\mu_i -\varepsilon_i)/T]\right\}
\eea
where $\epsilon_i(k)=(k^2+ m_i^2)^{1/2}$ is the $i-$th particle dispersion relation, $g_i$ is
spin-isospin degeneracy factor and $\pm$ stands for bosons and fermions, respectively.

Before the discovery of QCD, a probable phase transition of a massless pion gas to a new phase of matter was speculated \cite{lsm1}. Based on statistical models like Hagedorn \cite{hgdrn1} and Bootstrap \cite{boots1}, the thermodynamics of such an ideal pion gas is studied, extensively. After the QCD, the new phase of matter is known as quark gluon plasma (QGP). The physical picture was that at $T_c$ the additional degrees of freedom carried by QGP are to be released resulting in an increase in the thermodynamic quantities like energy and pressure densities. The success of hadron resonance gas model in reproducing lattice QCD results at various quark flavours and masses (below $T_c$) changed this physical picture drastically. Instead of releasing additional degrees of freedom at $T>T_c$, the interacting system reduces its effective degrees of freedom at $T<T_c$. In other word, the hadron gas has much more degrees of freedom than QGP.

At finite temperature $T$ and baryon chemical potential $\mu_i $, the pressure of the $i$-th hadron or resonance species reads 
\begin{equation}
\label{eq:prss} 
p(T,\mu_i ) = \pm \frac{g_i}{2\pi^2}T \int_{0}^{\infty}
           k^2 dk  \ln\left\{1 \pm \exp[(\mu_i -\varepsilon_i)/T]\right\}.
\end{equation} 
As no phase transition is conjectured in HRG, summing over all hadron resonances results in the final thermodynamic pressure in the hadronic phase. 

The switching between hadron and quark chemistry is given by the relations between  the {\it hadronic} chemical potentials and the quark constituents; 
$\mu_i =3\, n_b\, \mu_q + n_s\, \mu_S$, where $n_b$($n_s$) being baryon (strange) quantum number. The chemical potential assigned to the light quarks is $\mu_q=(\mu_u+\mu_d)/2$ and the one assigned to strange quark reads $\mu_S=\mu_q-\mu_s$. The strangeness chemical potential $\mu_S$ is
calculated as a function of $T$ and $\mu_i $ under the assumption that the overall
strange quantum number has to remain conserved in heavy-ion collisions~\cite{Tawfik:2004sw}.


\section{Normalized Higher Order Moments of Particle Multiplicity}
\label{sec:norm}

\begin{figure}[htb]
\includegraphics[angle=-90,width=12cm]{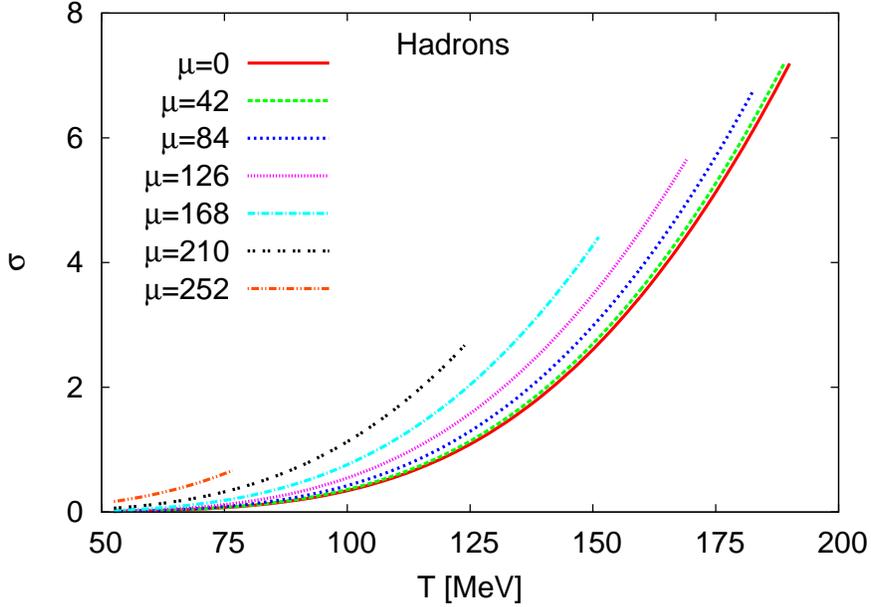}
\caption{Results for  $\sigma$ in hadronic  resonances are given in dependence on $T$ for various baryon chemical potentials (given in MeV).}
\label{fig:sigmaaa} 
\end{figure}

The normalization of higher order moments which can be deduced through derivatives of Eq. (\ref{eq:prss}) with respect to the chemical potential $\mu$ of given charges, apparently gives additional insights about the properties of higher order moments. From statistical point of view, the normalization is done with respect to the standard deviation $\sigma$, which is be related to $\xi$. Therefore, it provides with a tool to relate moments with various orders to the experimental measurement. The susceptibility of the distribution give a measure for $\sigma$.  For instance, the susceptibility is given by the derivative of first order moments with respect to $\mu$. It has been shown that the susceptibility is related to $\sim\xi^2$ \cite{endp1}. The results of  $\sigma$ in hadronic resonances are calculated at different $\mu$ and given in Fig. \ref{fig:sigmaaa}. As per the standard model, strangeness is one of the global symmetries in strong interactions. The procedure of keeping strange degrees of freedom conserved in HRG is introduced in Ref. \cite{Tawfik:2004sw}. Although, the baryon chemical potential $\mu$ vanishes per definition, the chemical potential associated with the strange quark $\mu_S$ remains finite.  At chemical freeze-out boundary, the dependence of $\sigma$ on $\mu$ is given in Fig. \ref{fig:S2n2}.

The normalization of $4-$th order moment is known as heteroskedacity or kurtosis. It means varying volatility or more accurately, varying variance.  Actually, the kurtosis is given by normalized $4-$th order moment minus $3$. The subtraction of $3$, which arises from the Gaussian distribution,  is usually omitted \cite{kurts1,kurts2,kurts3}.  Therefore, the kurtosis is an ideal quantity for probing the non-Gaussian fluctuation feature as expected near $T_c$ and critical endpoint. A sign change of skewness or kurtosis is conjectured to indicate that the system crosses a boundary that might change the symmetry \cite{endp5,endp6}. As HRG is valid below $T_c$, the sign change is not accessible. It has been shown that kurtosis $\kappa$ is related to $\sim\xi^{7}$ \cite{endp4}. The kurtosis in bosonic and fermionic resonances, respectively, reads
\bea
\kappa_b &=& -\frac{\pi^2}{g_i}\, T^3\, \frac{\int_0^{\infty}
  \left\{\text{cosh}\left[\frac{\varepsilon_i - \mu_i }{T}\right] + 2\right\} 
\text{csch}\left[\frac{\varepsilon_i - \mu_i }{2\, T}\right]^4 \; k^2 \, dk}
{\left[\int_0^{\infty} \left(1-\text{Cosh}\left[\frac{\varepsilon_i-\mu_i }{T}\right]\right)^{-1}\; k^2 \, dk\right]^{2} } - 3,  \label{eq:Kkb}\\
\kappa_f &=& \frac{\pi^2}{g_i}\, T^3\, \frac{\int_0^{\infty}
  \left\{\text{cosh}\left[\frac{\varepsilon_i - \mu_i }{T}\right] - 2\right\} 
\text{sech}\left[\frac{\varepsilon_i - \mu_i }{2\, T}\right]^4 \; k^2 \, dk}
{\left[\int_0^{\infty} \left(\text{cosh}\left[\frac{\varepsilon_i-\mu_i }{T}\right]+1\right)^{-1}\; k^2 \, dk\right]^{2} } - 3.  \label{eq:Kkf}
\eea


\subsection{Products of higher order moments}
\label{sec:mult}

There are several techniques to scale the correlation functions. The survey system's optional statistics module represents the most common technique i.e., Pearson or product moment correlation. This module includes the so-called partial correlation which seems to be useful when the relationship between two variables is to be highlighted, while effect of one or two other variables can be removed. In the present work, we study the  products of higher order moments of the distributions of conserved quantities. The justification of this step is that certain products can be directly connected to the corresponding susceptibilities as observed in lattice QCD simulations and related to the long range correlations \cite{endp5,qcdlike,latqcd1}. Seeking for simplicity, we start with the Boltzmann approximation.

When relativistic momentum integrals are replaced by summation over modified
Bessel functions, the $\sigma^2/\langle N\rangle$ and $\kappa\, \sigma^2$ in Boltzmann approximation, respectively,  reads
\bea 
\frac{\sigma^2}{\langle N\rangle} &=& \frac{\sum_{n=1}^{\infty}  (\pm)^{n+1}\, e^{n\frac{\mu_i}{T}}
  \left(n\frac{m_i}{T}\right)^2 n^{-2} K_2\left(n\frac{m_i}{T}\right)}{\sum_{n=1}^{\infty}  (\pm)^{n+1} e^{n\frac{\mu_i}{T}}
  \left(n\frac{m_i}{T}\right)^2 n^{-3} K_2\left(n\frac{m_i}{T}\right)}, \label{eq:k21} \\
\kappa\, \sigma^2 &=& \frac{\sum_{n=1}^{\infty} (\pm)^{n+1} e^{n\frac{\mu_i}{T}}
  \left(n\frac{m_i}{T}\right)^2 K_2\left(n\frac{m_i}{T}\right)}{\sum_{n=1}^{\infty} (\pm)^{n+1} e^{n\frac{\mu_i}{T}}
  \left(n\frac{m_i}{T}\right)^2 n^{-2} K_2\left(n\frac{m_i}{T}\right)} \nonumber \\
  &-& 3
\frac{g_i}{2 \pi^2} T^4 \sum_{n=1}^{\infty} (\pm)^{n+1}\,
e^{n\frac{\mu_i}{T}}\, \left(n\frac{m_i}{T}\right)^2 n^{-2}\, K_2\left(n\frac{m_i}{T}\right).  \label{eq:k23}
\eea
The origin of the second term in Eq. (\ref{eq:k23}) is obvious. Furthermore,  this equation apparently justifies the conclusions in \cite{KF2010} that in Boltzmann approximation 
\bea
\frac{\sigma^2}{\langle N\rangle} &=& S\, \sigma \simeq 1,
\eea
while
\bea \label{ea:kapsig1}
\kappa\, \sigma^2  &\simeq& 1 - 3 \frac{g_i}{2 \pi^2} T^4 \text{exp}\left[\frac{\mu_i}{T}\right]\, \left(\frac{m_i}{T}\right)^2 \, K_2\left(\frac{m_i}{T}\right).
\eea
These expressions are valid in the final state, which can be characterized by chemical and thermal freeze-out. In other words, they depend on the chemical potential. Using relativistic momentum integrals shows that the products of moments result in a constant dependence on $T$ \cite{KF2010}. 

The fluctuations of conserved quantities are assumed to be sensitive to the structure of the hadronic system in its final state. As mentioned above, crossing the phase boundary or passing through critical endpoint is associated with large fluctuations. Most proposed fluctuation of observables are variations of second order moments of the distribution, such as particle ratio \cite{Tawfik:2010pt,tawDF} and charged dynamical measurement \cite{chargeD}. Then, the fluctuations are approximately related to $\xi^2$ \cite{corrlenght}.

\begin{figure}[htb]
\includegraphics[angle=-90,width=12cm]{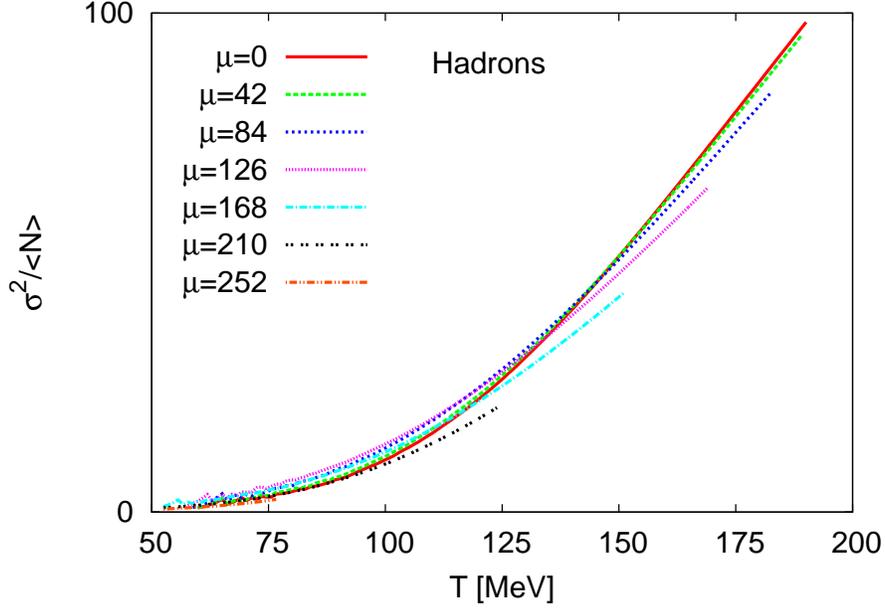}
\caption{The ratio $\sigma^2/\langle N\rangle$  is given in dependence on $T$ at different chemical potentials $\mu$ for hadronic  resonance gas. }
\label{fig:S2n} 
\end{figure}

\begin{figure}[htb]
\includegraphics[angle=-90,width=12.cm]{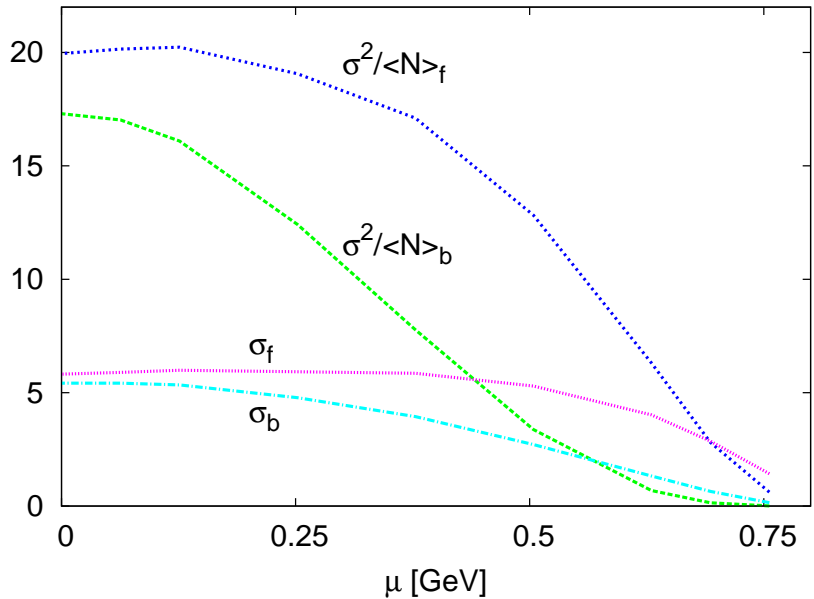}
\caption{The normalized moment $\sigma^2/\langle N\rangle$ and standard deviation $\sigma$ are given in dependence on $\mu$ for boson and fermion resonances. }
\label{fig:S2n2} 
\end{figure}

The ratio of standard deviation $\sigma^2$ and the mean multiplicity $\langle N \rangle$ for fermions and bosons reads
\bea
\frac{\sigma^2}{\langle N\rangle} &=& \frac{1}{2}\, \frac{\int_0^{\infty}
  \left(1\pm \text{csch} \left[\frac{\varepsilon_i-\mu_i}{T}\right]\right)^{-1}\, k^2\, dk}{\int_0^{\infty} \left(1 \pm e^{\frac{\varepsilon_i-\mu_i }{T}}\right)^{-1}\, k^2\, dk} \label{eq:sMb}, 
\eea
where $\pm$ stands for fermions and bosons, respectively. The results are given in Fig. \ref{fig:S2n}. We  notice that  the results are not spread. The thermal evolution of $\sigma^2/\langle N\rangle$ seems not depending on $\mu$. To illustrate the effect of $\mu$, we show in Fig. \ref{fig:S2n2} the normalized moment $\sigma^2/\langle N\rangle$ as a function of $\mu$ calculated at the chemical freeze-out boundary, which is characterized by $s/T^3=7$. We differentiate between fermions and bosons. We compare the results with the standard deviation $\sigma$. At large chemical potential, both quantities are almost equal. Decreasing $\mu$ increases the difference between $\sigma^2/\langle N\rangle$ and $\sigma$.

\begin{figure}[htb]
\includegraphics[angle=-90,width=12cm]{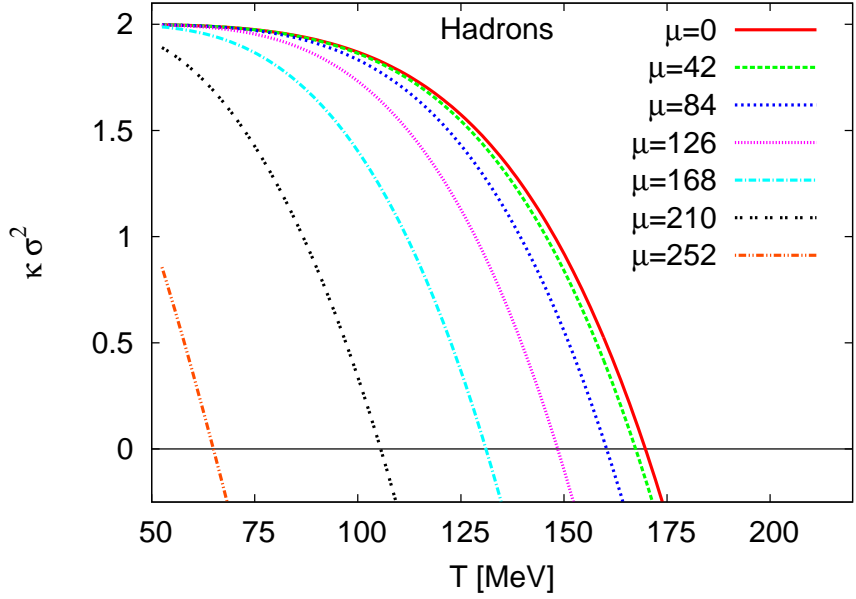}
\caption{The product ${\kappa}\,\sigma^2$ is given as a function of $T$ at various $\mu$-values as calculated in the HRG model. }
\label{fig:kS2muu} 
\end{figure}

The multiplication of kurtosis by $\sigma^2$ called $\kappa^{eff}$ \cite{keff} is apparently equivalent to the ratio of $3$-rd order moment to $2$-nd order moment.  In lattice QCD  and QCD-like models, $\kappa^{eff}$ is found to diverge near the critical endpoint \cite{endp5,qcdlike}. In HRG, the bosonic and fermionic products read
\bea 
\left(\kappa\; \sigma^2\right)_b &=& -\frac{1}{4} \frac{\int_0^{\infty}  \left\{\text{cosh}\left[\frac{\varepsilon_i
      -\mu_i }{T}\right] + 2\right\}\; \text{csch}\left[\frac{\varepsilon_i
      -\mu_i }{2\, T}\right]^4 \; k^2\,dk} {\int_0^{\infty} \left(1-\text{cosh}
  \left[\frac{\varepsilon_i -\mu_i }{T}\right]\right)^{-1}\; k^2\,dk} \nonumber \\
&+& \frac{3\, g_i}{4\, \pi^2}\, \frac{1}{T^3}\, \int_0^{\infty} \left(1-\text{cosh}
  \left[\frac{\varepsilon_i -\mu_i }{T}\right]\right)^{-1}\; k^2\,dk, \hspace*{7mm}\label{eq:lsigma2b} \\
\left(\kappa\; \sigma^2\right)_f &=& \frac{1}{4} \frac{\int_0^{\infty}  \left\{\text{cosh}\left[\frac{\varepsilon_i
      -\mu_i }{T}\right] - 2\right\}\; \text{Sech}\left[\frac{\varepsilon_i
      -\mu_i }{2\, T}\right]^4 \; k^2\,dk} {\int_0^{\infty} \left(\text{cosh}
  \left[\frac{\varepsilon_i -\mu_i }{T}\right]+1\right)^{-1}\; k^2\,dk} \nonumber \\
&-& \frac{3\, g_i}{4\, \pi^2}\, \frac{1}{T^3}\, \int_0^{\infty} \left(\text{cosh}
  \left[\frac{\varepsilon_i -\mu_i }{T}\right]+1\right)^{-1}\; k^2\,dk. \label{eq:lsigma2f}
\eea
When ignoring the constant term in Eqs. (\ref{eq:Kkb}) and (\ref{eq:Kkf}), then the second terms in the previous expressions entirely disappear. The results are given in Fig. \ref{fig:kS2muu}. In the hadronic sector, the dependence of $\kappa\,\sigma^2$ on the temperature $T$ at different $\mu$-values is depicted. We notice that increasing $T$ is accompanied with a drastic declination in $\kappa\,\sigma^2$. Also, we find that ${\kappa}\,\sigma^2$ flips its sign at large $T$. Also when comparing their dependences on the chemical potentials at the freeze-out boundary, Fig. \ref{fig:fezeout-sT3}, it is apparent that the $\mu$-dependence increases when the normalized fourth order moment is included. The product $\kappa\, \sigma^2$ calculated at the freeze-out boundary leads to some interesting findings. First, $\kappa\, \sigma^2$ almost vanishes or even flips its sign. Second, the $T$ and $\mu$ corresponding to vanishing $\kappa\, \sigma^2$ are coincident with the phenomenologically measured freeze-out parameters.

\section{Chemical Freeze-out and Normalized Higher Order Moments}
\label{sec:chemFO}

In rest frame of produced particle, the hadronic matter can be determined by constant degrees of freedom, for instance, $s/T^3 (4/\pi^2)=const$ \cite{Tawfik:2005qn,Tawfik:2004ss}. The quantity $const$ is assigned to $5$ and $7$ for two and three quark flavors, respectively. The chemical freeze-out is related to the particle creation. Therefore, the abundances of different particle species are controlled by the chemical potential, which obviously depends on $T$. With the beam energy, $T$ is increasing, while the baryon densities at mid-rapidity is decreasing. The estimation of the macroscopic parameters of the chemical freeze-out can be extracted from particle ratio. These parameters collected over the last three decades seem to fellow regular patterns as the beam energy increases \cite{jean2006,Tawfik:2005qn,Tawfik:2004ss}. The higher order moments have been suggested to control the chemical freeze-out, so that several conditions have been proposed \cite{HM_FO}.

As introduced in section \ref{sec:mult}, the thermal evolution of ${\kappa}\,\sigma^2$ slowly decreases. It vanishes and changes its sign. This result seems to update previous studies \cite{lqcd1a,r42}, where ${\kappa}\,\sigma^2$ was assumed to remain finite and positive with increasing $\mu$. In the present work, we find that the sign of ${\kappa}\,\sigma^2$ is flipped at high $T$ \cite{r43}. Furthermore, we find that the $T$- and $\mu$-parameters, at which the sign is flipped are amazingly coincide with the ones of the chemical freeze-out. 
Vanishing $\kappa\, \sigma^2$ for boson and fermion, respective, reads
\bea 
\int_0^{\infty}  \left\{\text{cosh}\left[\frac{\varepsilon_i
      -\mu_i }{T}\right] + 2\right\}\; \text{csch}\left[\frac{\varepsilon_i
      -\mu_i }{2\, T}\right]^4 \; k^2\,dk 
&=& \frac{3\, g_i}{\pi^2}\, \frac{1}{T^3}\, \left[\int_0^{\infty} \left(1-\text{cosh}
  \left[\frac{\varepsilon_i -\mu_i }{T}\right]\right)^{-1}\; k^2\,dk\right]^2, \hspace*{7mm}\label{eq:lsigma2g} \\
\int_0^{\infty}  \left\{\text{cosh}\left[\frac{\varepsilon_i
      -\mu_i }{T}\right] - 2\right\}\; \text{sech}\left[\frac{\varepsilon_i
      -\mu_i }{2\, T}\right]^4 \; k^2\,dk 
&=& \frac{3\, g_i}{\pi^2}\, \frac{1}{T^3}\, \left[\int_0^{\infty} \left(\text{cosh}
  \left[\frac{\varepsilon_i -\mu_i }{T}\right]+1\right)^{-1}\; k^2\,dk\right]^2. \hspace*{10mm}\label{eq:lsigma2fc}
\eea
The rhs and lhs in both expressions can be re-written as 
\bea
16  \frac{\pi^2}{g_i} T^3\, m_4(T,\mu) &=& 48 \frac{\pi^2}{g_i} T^3\, \chi^2(T,\mu),
\eea
which is valid for both bosons and fermions. Then, the chemical freeze-out is defined, if the condition
\bea
m_4(T,\mu) &=& 3\, \chi^2(T,\mu),
\eea 
is fulfilled. At the chemical freeze-out curve, a naive estimation leads to $\xi\sim 3^{1/3}~$fm. In doing this, it is assumed that the proportionality coefficients of  $\kappa \sim \xi^7$ and $\chi\sim\xi^2$, are equal.  An estimation for $\xi$  in the heavy-ion collisions has been reported \cite{xxii2}. Near a critical point, the experimental value $\sim 2-3\,$fm (only factor 3 larger) agrees well with our estimation.

\begin{figure}[htb]
\includegraphics[angle=-90,width=12.cm]{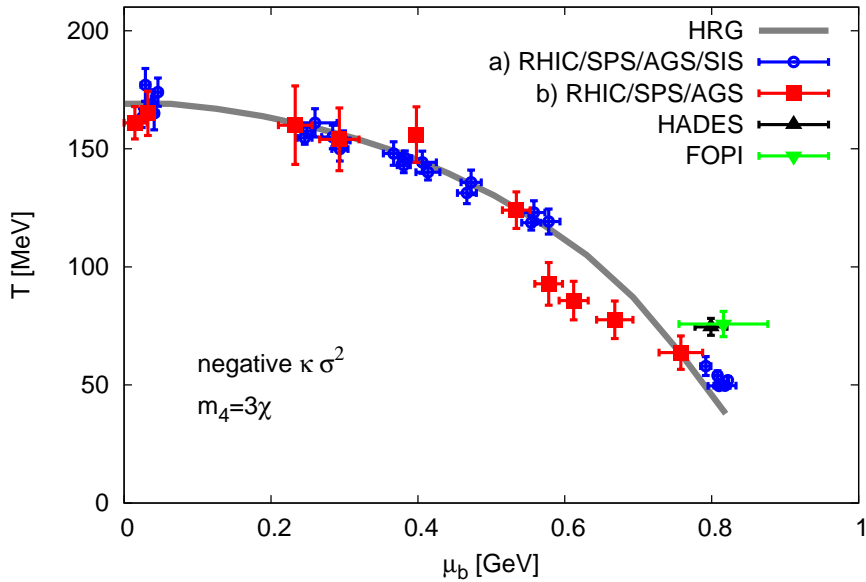}
\caption{The chemical freeze-out boundary is given in dependence on the temperature $T$. The experimental data are presented by the solid circles. The curves represent the results of HRG determined when the sign of ${\kappa}\, \sigma^2$ is flipped. The filled circles are taken from \cite{jean2006} and filled squares from \cite{dataCR}. The upwards and downwards triangle represent HADES \cite{hades} and FOPI \cite{fopi} results, respectively. }
\label{fig:fezeout-sT3} 
\end{figure}

At the chemical freeze-out curve, the intensive parameters $T$ and $\mu$ which are related to the extensive properties entropy and particle number, respectively, have to be determined over a wide range of beam energies. Fig. \ref{fig:fezeout-sT3} collects a large experimental data set. For a recent review, we refer to \cite{jean2006} (filled squares) and the references therein. The filled circles are taken from \cite{dataCR}. The upwards and downwards triangle represent HADES \cite{hades} and FOPI \cite{fopi} results, respectively. The solid curve represents a set of $T$ and $\mu$, at which ${\kappa}\,\sigma^2$ vanishes as calculated in HRG. It is obvious that this curve reproduces very well the experimental data. As given above, at this curve the normalized fourth order moment $\kappa$ is equal to three times the squared susceptibility $\chi$.  This new condition seems to guarantee the condition introduced in \cite{Tawfik:2005qn,Tawfik:2004ss}; $s/T^3=const.$ over the range $0<\mu<0.8~$GeV.

\subsection{Physics of vanishing $\kappa\, \sigma^2$ or equivalently $m_4=3\chi^2$}

The kurtosis is assumed to give the degree of peakedness of a given distribution and the heaviness of its tail. Alternatively, the kurtosis is defined as the location- and scale-free movement of the highest probability from the shoulders of a distribution to its center and tails \cite{kurts}. In the present work, the kurtosis ''excess'' is utilized, for instance Eqs. (\ref{eq:Kkb}) and (\ref{eq:Kkf}) for fermions and bosons, respectively. The kurtosis excess belongs to a normal distribution that vanishes. Whether kurtosis or kurtosis ''excess'', what matters is the origin or the nature of the variance $\sigma$. If the data is not peaky, it is said that the variance is distributed throughout the distribution. But if the data is peaky, then the variance is supposed to go closer to the ''center''. In that case, its origin is the tail.

The variance is related to the second order momentum, the susceptibility, $\sigma^2=\chi$ \cite{Tawfik:2012si}. Furthermore, it is found that the product of normalized moments $\kappa\, \sigma^2$ is equivalent to the ratio of non-normalized quartic to quadratic order moments $m_4/m_2$. On one hand, the fluctuations of $m_4/m_2$ of the net quark number, $R_{4,2}=m_4/m_2$, is studied in the lattice QCD at finite $T$ and $\mu$ \cite{latticeQCD1a,fodorFO} and interpreted as a valuable probe of deconfinement and chiral dynamics \cite{latticeQCD1a}. Furthermore, the inverse compressibility, $R_{\kappa}$,  calculated in some effective models is interpreted as a useful observable for identifying the position of the critical endpoint in the QCD phase diagram \cite{modelll}. On the other hand, comparing the ratios of higher order moments of electric charge to preliminary data on higher order moments of the net electric charge distribution from the STAR collaboration, it is concluded that freeze-out temperature and chemical potential can be determined from first principles \cite{fodorFO}. It is worthwhile to mention that the connections between the ratios of higher order moments and the freeze-out parameters lie on the same line with the present and previous work \cite{Tawfik:2012si}. 

Accordingly, the freeze-out parameters are given, if $\kappa\, \sigma^2$ vanishes or equivalently $m_4/m_2=3$ i.e., 
\bea
\frac{\int_0^{\infty} k^2\, dk \left[\cosh(\frac{\varepsilon-\mu}{T})\pm2\right] \left[\exp(\frac{\varepsilon-\mu}{2T})\mp\exp(\frac{\mu-\varepsilon}{2T})\right]^4}{\int_0^{\infty} k^2\, dk \left[\exp(\frac{\varepsilon-\mu}{2T})\mp\exp(\frac{\mu-\varepsilon}{2T})\right]^2} &=&\frac{3}{4}.
\eea
The dependence of $\kappa\,\sigma^2$ on the temperature $T$  is presented in Fig. \ref{fig:kS2muu} at different chemical potentials $\mu$. We find  a drastic decrease in $\kappa\,\sigma^2$ with increasing $T$. Furthermore, the decrease continues so that at a certain value of $T$, the sign of ${\kappa}\,\sigma^2$ flips out. The temperature at which ${\kappa}\,\sigma^2$-sign flips depends on $\mu$. The resulting $T$ and $\mu$ are coincident with the phenomenologically measured freeze-out parameters, Fig. \ref{fig:fezeout-sT3}.

\subsection{Other Conditions for Chemical Freeze-Out Parameters}

Starting from phenomenological observations at SIS energy, it was found that the averaged energy per averaged particle $ \epsilon /  n \approx 1~$GeV \cite{jeanRedlich}, where Boltzmann approximations are applied in calculating  $ \epsilon /  n $, this constant ratio is assumed to describe the whole $T-\mu_b$ diagram. For completeness, we mention that the authors assumed that the pions and rho-mesons get dominant, at high $T$ and small $\mu_b$. The second criterion assumes that total baryon number density $  n_b +  n_{\bar{b}} \approx 0.12~$fm$^{-3}$ \cite{nb01}. In framework of percolation theory, the authors of Ref. \cite{percl} have suggested a third criterion. As shown in Fig. 2 of \cite{Tawfik:2005qn}, the last two criteria seem to give almost identical results. All of them are stemming from phenomenological observation. A fourth criterion based on lattice QCD simulations was introduced in Ref.  \cite{Tawfik:2005qn,Tawfik:2004ss}. Accordingly, the entropy normalized to cubic temperature is assumed to remain constant over the whole range of baryo-chemical potentials, which is related to the nucleus-nucleus center-of-mass energies $\sqrt{s_{NN}}$ \cite{jean2006}. An extensive comparison between constant $ \epsilon /  n $ and constant $s/T^3$ is given in \cite{Tawfik:2005qn,Tawfik:2004ss}. 

In framework of hadron resonance gas, the thermodynamic quantities deriving the chemical freeze-out are deduced \cite{Tawfik:2005qn,Tawfik:2004ss}. The motivation of suggesting constant normalized entropy is the comparison to the lattice QCD simulations with two and three flavors. We simply found the $s/T^3=5$ for two flavors and $s/T^3=7$ for three flavors. Furthermore, we confront the hadron resonance gas results to the experimental estimation for the freeze-out parameters, $T$ and $\mu_b$. 

In the present work, we introduce a novel condition characterizing the freeze-out parameters. To this extend, the higher order moments are applied \cite{Tawfik:2012si}. Vanishing ${\kappa}\, \sigma^2$ or equivalently $m_4/\chi=3$ results in $T$ - $\mu$ sets coincident with the phenomenologically estimated ones.  Recently, lattice QCD calculations confirm the same connection between the ratios of higher order fluctuations and the freeze-out parameters \cite{fodorFO,nakamura1}. 

\section{Conclusion}
\label{sec:conc}

In the present work, the non-normalized order moments of the particle multiplicity are calculated in the HRG model. We studied the thermal evolution of the first four normalized order moments and their products (ratios) at different chemical potentials $\mu$. By doing that, the evaluate the normalized moments are estimated at the chemical freeze-out curve.  It has been found  that non-monotonic behavior reflecting dynamical fluctuation and strong correlations appears starting from the normalized third order moment (skewness $S$). 
Furthermore, non-monotonicity is observed in the normalized fourth order moment, the kurtosis $\kappa$, and its products. These are novel observations. Although HRG is exclusively applicable below $T_c$ i.e. it does not include deconfinement phase transition, it is apparent that the higher order moments are able to give signatures for the critical phenomena. Based on these findings, we introduced novel conditions characterizing the chemical freeze-out curve as follows. The chemical freeze-out curve is described by $m_4=3\,\chi^2$, where $\chi$ is the susceptibility in particle number, i.e. the second order moment. We are able to estimate the freeze-out parameters, although the hadron resonance gas model basically does not contain information on the criticality related with the chiral dynamics and singularity in the physical observables required to locate the critical endpoint, for instance. After submitting this work, a new paper was posted to {\tt arXiv}, in which the authors used the second order moment (susceptibility) and the fourth order one (kurtosis) as they are apparently sensitive to the phase transition \cite{fodorFO,nakamura1}. Furthermore, comparing the ratios of higher order moments of electric charge to preliminary data on higher order moments of the net electric charge distribution from the STAR collaboration, it was concluded that freeze-out parameters can be determined from first principles (lattice QCD) \cite{fodorFO}.






\begin{thebibliography}{99}

\bibitem{Tawfik:2012si} A. Tawfik, to appear in Adv. High Energy Phys., arXiv:1205.1761 [hep-ph]

\bibitem{endp1} C. Athanasiou, K. Rajagopal and M. Stephanov,  Phys. Rev. D {\bf 82}, 074008 (2010).

\bibitem{star1} Xiaofeng Luo [STAR Collaboration], Acta Phys. Polon. Supp. {\bf 5}, 497 (2012); J. Phys. Conf. Ser. {\bf 316}, 012003 (2011); M.M. Aggarwal {\it et al.} [STAR Collaboration], Phys. Rev. Lett. {\bf 105}, 022302 (2010). 

\bibitem{star2} T. J. Tarnowsky [STAR Collaboration], J. Phys. G {\bf 38}, 124054 (2011); T. K. Nayak [STAR Collaboration]  Nucl. Phys. A {\bf 830}, 555C-558C (2009).

\bibitem{goore} M. A. York, G. D. Moore, 1106.2535 [hep-lat].

\bibitem{endp4} M. A. Stephanov, Phys. Rev. Lett. {\bf 102}, 032301 (2009).

\bibitem{endp4a} Y. Zhou, {\it et al.}, Phys. Rev. C {\bf 82}, 014905 (2010).

\bibitem{corrlenght} M. Stephanov, K. Rajagopal and E. Shuryak, Phys. Rev. D {\bf 60}, 114028 (1999).

\bibitem{endp2} J. Adams {\it et al.}, Nucl. Phys. A {\bf 757}, 102 (2005).

\bibitem{Tawfik:2004sw} A.~Tawfik,~Phys.~Rev.~D {\bf 71}~054502~(2005).

\bibitem{jean2006} J. Cleymans, H. Oeschler, K. Redlich and S. Wheaton, Phys. Rev. C {\bf 73}, 034905 (2006).

\bibitem{Karsch:2003vd} F.~Karsch, K.~Redlich and A.~Tawfik,~Eur.~Phys.~J.~C {\bf 29},~549~(2003).

\bibitem{Karsch:2003zq} F.~Karsch, K.~Redlich and A.~Tawfik, Phys.~Lett.~B {\bf 571},~67~(2003).

\bibitem{Redlich:2004gp} K.~Redlich, F.~Karsch and A.~Tawfik, J.~Phys.~G {\bf 30},~S1271~(2004). 

\bibitem{Tawfik:2004vv} A. Tawfik, J. Phys. G {\bf G31}, S1105-S1110 (2005). 

\bibitem{Tawfik:2006yq} A. Tawfik, Indian J. Phys. {\bf 85}, 755-766 (2011).

\bibitem{Tawfik:2010uh} A. Tawfik,  Prog. Theor. Phys. {\bf 126}, 279-292 (2011). 

\bibitem{Tawfik:2010pt} A. Tawfik, Nucl. Phys. A {\bf 859}, 63-72 (2011).

\bibitem{Tawfik:2012zz} A. Tawfik,  Int. J. Theor. Phys. {\bf 51}, 1396-1407 (2012).

\bibitem{Tawfik:2005qn} A. Tawfik, Nucl. Phys. A {\bf 764}, 387-392 (2006).

\bibitem{Tawfik:2004ss} A. Tawfik, Europhys. Lett. {\bf 75}, 420 (2006).

\bibitem{Hagedorn:1965st} R. Hagedorn, Nuovo Cim. Suppl. {\bf 3}, 147 (1965).

\bibitem{Vunog} R.~Venugopalan, M.~Prakash, Nucl.~Phys.~A {\bf 546},~718~(1992). 

\bibitem{lsm1} M. Gell-Mann and M. Levy, 
Il Nuovo Cimento {\bf 16}, 705–726 (1960), doi:10.1007/BF02859738

\bibitem{hgdrn1} R. Hagedorn, Nuovo Cim. Suppl. {\bf 6}, 311-354 (1968); Nuovo Cim. A {\bf 56}, 1027-1057 (1968). 

\bibitem{boots1} R. J. Eden, P. V. Landshoff, D. I. Olive and J. C. Polkinghorne, The Analytic S-Matrix, Cambridge University Press, 1966; 
J. Letessier‏, J. Rafelski‏, Hadrons and quark-gluon plasma, Cambridge University Press, 2002. 

\bibitem{kurts1} K. P. Ballanda and H. L. MacGillivray, ''Kurtosis: A Critical Review'', The American Statistician, {\bf 42}, 111 - 119 (1988).

\bibitem{kurts2} N. L. Johnson, S. Kotz and N. Balakkrishan, ''Continuous Univariate Distribution'', Wiley-Interscience Publication, Houghton Mifflin, 1970.

\bibitem{kurts3} C. T. Hsu and D. N. Lawley, Biometrika {\bf 31}, 238-248 (1940).

\bibitem{endp5} M. Cheng {\it et al.}, Phys. Rev. D {\bf 79}, 074505 (2009).

\bibitem{endp6}  M. Kitazawa, M. Asakawa and S. Ejiri, Phys. Rev. Lett. {\bf 103}, 262301 (2009); PoS LAT2009, 174 (2009).

\bibitem{qcdlike} B. Stokic {\it et al.}, Phys. Lett. B, {\bf 673}, 192 (2009). 

\bibitem{latqcd1} R. V. Gavai and S. Gupta, Phys. Lett. B {\bf 696}, 459 (2011).

\bibitem{KF2010} F. Katsch and K. Redlich, Phys. Lett. B {\bf 695}, 136-142 (2010).

\bibitem{tawDF} A. Tawfik, 
J. Phys. G {\bf 40}, 055109 (2013);   
Indian J. Phys. {\bf 86}, 1139-1146 (2012) ;  
Indian J. Phys. {\bf 86}, 641-646  (2012); 
hep-ph/0604037; 
M.I. Adamovich {\it et al.} [EMU01 Collaboration], Heavy Ion Phys. {\bf 13}, 213-221 (2001).

\bibitem{chargeD}  B. I. Abelev {\it et al.}, [STAR Collaboration], Phys. Rev. C {\bf 79}, 024906 (2009).  

\bibitem{keff} T. Schuster, M. Nahrgang, M. Mitrovski, R. Stock and  M. Bleicher, Eur. Phys. J. C {\bf 72},  2143  (2012).


\bibitem{HM_FO} 
B. Friman, F. Karsch, K. Redlich and V. Skokov, Eur. Phys. J. C {\bf 71}, 1694 (2011); 
F. Karsch and K. Redlich, Phys. Lett. B {\bf 695}, 136-142 (2011); 
R. V. Gavai and Sourendu Gupta, Phys. Lett. B {\bf 696}, 459-463 (2011); 
F. Karsch, J. Phys. G {\bf 38}, 124098 (2011).

\bibitem{lqcd1a} C. Miao [RBC-Bielefeld Collaboration],  Nucl. Phys. A {\bf 830}, 705C-708C (2009). 

\bibitem{r42} F. Karsch, B.-J. Schaefer, M. Wagner, and J. Wambach, Phys. Lett. B {\bf 698}, 256-264 (2011);  F. Karsch and K. Redlich, Phys. Lett. B {\bf 695}, 136-142 (2011).

\bibitem{r43} B. Friman, F. Karsch, K. Redlich, and V. Skokov,  Eur. Phys. J. C {\bf 71}, 1694  (2011).

\bibitem{xxii2} B. Berdnikov {\it et al.}, Phys. Rev. D {\bf 61}, 105017 (2000).

\bibitem{dataCR} A. Andronic, P. Braun-Munzinger and J. Stachel, Nucl. Phys. A {\bf 772}, 167 (2006).


\bibitem{hades}  G. Agakishiev {\it et al.} [HADES Collaboration], Eur. Phys. J. A {\bf 47},  21 (2011).

\bibitem{fopi} X. Lopez {\it et al.} [FOPI Collaboration], Phys. Rev. C {\bf 76}, 052203 (2007).

\bibitem{kurts} K. P. Balanda and H.L. MacGillivray, 
The American Statistician, {\bf 42}, 111-119  (1988).

\bibitem{latticeQCD1a} C. R. Allton, M. Doring, S. Ejiri, S.J. Hands, O. Kaczmarek, F. Karsch, E. Laermann and K. Redlich, Phys. Rev. D {\bf 71},
054508 (2005); S. Ejiri, F. Karsch and K. Redlich, Phys. Lett. B {\bf 633}, 275 (2006).

\bibitem{fodorFO} S. Borsanyi, Z. Fodor, S. D. Katz, S. Krieg, C. Ratti and K. K. Szabo, arXiv:1305.5161 [hep-lat]

\bibitem{modelll} C. Sasaki, B. Friman and K. Redlich, Phys. Rev. Lett. {\bf 99}, 232301 (2007).

\bibitem{jeanRedlich} J. Cleymans and K. Redlich, Phys. Rev. C {\bf 60}, 054908 (1999).

\bibitem{nb01} P. Braun-Munzinger and J. Stachel, J. Phys. G {\bf 28}, 1971 (2002).

\bibitem{percl} V. Magas and H. Satz, Eur. Phys. J. C {\bf 32}, 115 (2003).

\bibitem{nakamura1} A. Nakamura and K. Nagata, arXiv:1305.0760 [hep-ph]


\end{thebibliography}
\end{document}